\documentclass[aps,nofootinbib,twocolumn]{revtex4}
\pdfoutput=1
\usepackage{amsfonts,amssymb,amsmath}            
\usepackage[]{graphics,graphicx,epsfig}            
\usepackage{amsthm}
\usepackage{hyperref}
\def\identity{\leavevmode\hbox{\small1\kern-3.8pt\normalsize1}}

\newcommand{\ket}[1]{\left | #1 \right\rangle}
\newcommand{\bra}[1]{\left \langle #1 \right |}
\newcommand{\half}{\mbox{$\textstyle \frac{1}{2}$}}

\newcommand{\braket}[2]{\left\langle #1|#2\right\rangle}
\newcommand{\proj}[1]{\ket{#1}\bra{#1}}
\newcommand{\Tr}{\text{Tr}}
\renewcommand{\epsilon}{\varepsilon}
\begin{document}

\title{The Basics of Perfect Communication through Quantum Networks}
\date{\today}

\author{Alastair \surname{Kay}}
\affiliation{Centre for Quantum Technologies, National University of Singapore, 
			3 Science Drive 2, Singapore 117543}
\affiliation{Keble College, Parks Road, Oxford, OX1 3PG}
            
\begin{abstract}
Perfect transfer of a quantum state through a one-dimensional chain is now well understood, allowing one not only to decide whether a fixed Hamiltonian achieves perfect transfer, but to design a suitable one. We are particularly interested in being able to design, or understand the limitations imposed upon, Hamiltonians subject to various naturally arising constraints such as a limited coupling topology with low connectivity (specified by a graph) and type of interaction. In this paper, we characterise the necessary and sufficient conditions for transfer through a network, and describe some natural consequences such as the impossibility of routing between many different recipients for a large class of Hamiltonians, and the limitations on transfer rate. We also consider some of the trade-offs that arise in uniformly coupled networks (both Heisenberg and $XX$ models) between transfer distance and the size of the network as a consequence of the derived conditions.
\end{abstract}

\maketitle

The task of state transfer was introduced \cite{bos03} with the intention of reducing the control required to communicate between distant qubits in a quantum computer. The perfect action is defined by starting with a single qubit state $\rho_{\text{qubit}}$ on some input node, $A$, and $\rho_{\text{in}}$ is the state of the rest of the many-qubit system, and requiring that after evolution for some time $t_0$ under a fixed Hamiltonian $H$, the output state
$$
e^{-iHt_0}(\rho^A_{\text{qubit}}\otimes\rho_{\text{in}})e^{iHt_0}=\rho^B_{\text{qubit}}\otimes\rho_{\text{out}}
$$
is produced, thereby transmitting the input qubit to another site, $B$. Early attempts concentrated on a one-dimensional geometry. Transferring states perfectly by such a scheme requires the precise tuning of coupling strengths \cite{mcprl,lambrop,mcpra}, and the necessary and sufficient conditions for the transfer are now well understood \cite{akreview}, allowing us both to recognise and design \cite{stolze} perfect transfer chains. There are, however, a number of limitations that one might like to overcome. In the ideal case, we would have a uniformly coupled system which perfectly achieves high rates of transfer over large distances. It would have the ability to route states to different recipients (which is an important feature in increasing the connectivity within a quantum computer), and it would be less susceptible to static defects (Anderson localisation) \cite{keating} or dynamical fluctuations \cite{osborne_defects} than the one dimensional systems.

It has been proposed that spin networks could enable all of these properties, and certainly by allowing some small degree of control, they do \cite{pjprprl}. It is expected that in higher dimensional geometries, the effects of Anderson localisation are diminished. Also, if one allows a completely connected network, routing is possible between all the different nodes\footnote{One can simply define a permutation matrix that permutes through all the vertices as a target unitary, and fix the system Hamiltonian to be its logarithm.} \cite{nik}. Equally, perfect transfer is possible over long distances in a uniformly coupled network (such as the hypercube, \cite{mcpra}), although the best known scaling of the transfer distance with the number of vertices, $N$, is only $O(\log N)$. Unfortunately, while we know that networks exhibit a lot of potential, our level of understanding is still far from that of chains in spite of efforts by Godsil, Severini and coauthors in proving some necessary conditions \cite{sev1,sev2,godsil}. The purpose of this paper is to redress the balance by deriving necessary and sufficient conditions which will allow us to readily recognise perfect transfer networks i.e.\ the setting that we envisage is an experimentalist with a range of Hamiltonians available to them that they can implement, and they want to know if perfect transfer can be realised.

We can try and predict what some of the properties of such a Hamiltonian might be. For instance, the whole point of introducing state transfer \cite{bos03} was to reduce the difficulty of interacting distant sites with no direct coupling (for which transfer is trivial), so we impose the fact that there is likely to be an underlying geometry of feasible couplings, and that the type of these couplings is likely to be limited. As a consequence, we prove the impossibility of routing within the single excitation subspace of a broad class of systems (those whose Hamiltonians are real), and bound the maximum transfer rate for excitation preserving Hamiltonians. For uniformly coupled networks, we give the first upper bound on the transfer distance of a graph as a function of the number of vertices.

While the task of state transfer was proposed as a solution to a practical problem in quantum computers, it has provided a powerful technique for constructing other Hamiltonian-driven evolutions, and has been used to understand adversarial Hamiltonian perturbations \cite{aktorus}, quantum computation \cite{peres} etc. By moving beyond the study of chains to more general networks, we may be able to further extend the utility of these constructions.

\section{The Model}

State transfer has been almost universally studied for Hamiltonians $H$ which have the very particular property of excitation preservation,
$$
\left[H,\sum_{n=1}^NZ_n\right],
$$
which means that the number of qubits in the $\ket{1}$ state are a constant of the motion. If $k$ of the $N$ qubits are in the $\ket{1}$ state, then we say the system is in the $k^{th}$ excitation subspace\footnote{Those schemes that have been proposed which don't preserve the number of excitations have some unitarily equivalent conserved quantity \cite{akprl} or very similar symmetry \cite{paternostro,akreview}.}. These studies then proceed by setting the initial state of every qubit not $A$ to $\ket{0}$ so that when a state is placed on qubit $A$, we are in a superposition of the 0 and 1 excitation subspaces, and the task becomes
$$
(\alpha\ket{0}+\beta\ket{1})_A\ket{0}^{\otimes (N-1)}\mapsto(\alpha\ket{0}+e^{i\phi}\beta\ket{1})_B\ket{0}^{\otimes (N-1)}
$$
up to some phase $\phi$ which we could compensate for later (here we distinguish between {\em perfect} transfer and arbitrarily accurate transfer, concentrating on the former). The 0 excitation subspace is composed of a single state, and therefore does not evolve in time. Hence, we only have to concentrate on the evolution of the 1 excitation subspace, which is simply described by an $N\times N$ matrix, $H_1$. There are many different Hamiltonians that have the same $H_1$ such as the $XX$, Heisenberg and coupled harmonic oscillator Hamiltonians \cite{plenio}, not to mention more exotic varieties involving q-deformed oscillators \cite{qdef}.


We will continue to study transfer in the single excitation subspace of an excitation preserving Hamiltonian. For coupled harmonic oscillators (i.e.\ non-interacting bosons), this is no restriction since the single excitation subspace describes the dynamics of each individual boson no matter what other bosons are present in the system. While this is not true for spin systems\footnote{The Jordan-Wigner transformation maps 1D systems with the $XX$ Hamiltonian to non-interacting fermions, for which a similar argument holds \cite{albanese}.}, it is a reasonable restriction in the sense that higher excitation subspaces can be described as a single excitation subspace of a system of more qubits \cite{osborne}. The trade-off is that the necessary and sufficient conditions that we derive here are just necessary conditions in higher excitation subspaces for a given initial state on the rest of the system (with sufficiency assured if transfer of excitations is possible from the same initial state on two neighbouring excitation subspaces in the same time).

Within the single excitation subspace, we denote an excitation on qubit $n$ by
$$
\ket{n}:=\ket{0}^{\otimes n-1}\ket{1}\ket{0}^{\otimes (N-n)}.
$$
Consequently, the Hamiltonian is expressed as
$$
H_1=\sum_{n<m}J_{nm}\ket{n}\bra{m}+J_{nm}^*\ket{m}\bra{n}+\sum_{n=1}^NB_n\proj{n}.
$$
There are two physically important restrictions that can apply to the Hamiltonian. The first is to make the Hamiltonian real, i.e. $J_{nm}^*=J_{nm}$. It is not impossible that a Hamiltonian would contain complex coefficients, but most naturally arising, or readily implemented, Hamiltonian terms, such as $J(XX+YY)+\Delta ZZ$ only give rise to real Hamiltonians\footnote{Of course, this assumption readily generalises to Hamiltonians which, under an arbitrary local unitary transformation on each qubit, are real. In many scenarios, a local unitary transformation from a real Hamiltonian might typically be the way that complex coupling coefficients are generated \cite{pac3}.}. The second is a geometry constraint. Consider a graph $G$ which is composed of edges $E$ and vertices $V$, where we establish a correspondence between the $N$ vertices and the $N$ states $\ket{n}$. The only allowed non-zero couplings $J_{nm}$ are those for which the pair of vertices $n,m$ are an edge of the graph\footnote{Some works choose to set $J_{nm}=1$ for all edges, giving a correspondence between the adjacency matrix of the graph and $H_1$. While just a special case of our more general formalism, we will also discuss this restriction in Sec.\ \ref{sec:uniform}.}. The graph naturally induces a concept of transfer distance -- the minimum number of edges that one must follow to traverse the graph from the input node $A$ to the output node $B$. We consider short transfer distances uninteresting because they do not simplify the communication demands between distant qubits.

In our analysis, it will often help to work in the diagonal basis of $H_1$,
$$
H_1=\sum_{n=1}^N\lambda_n\proj{\lambda_n},
$$
i.e.\ $\ket{\lambda_n}$ is an eigenvector of $H_1$ of eigenvalue $\lambda_n$. Note that for real Hamiltonians, all the elements $\braket{m}{\lambda_n}$ must also be real.

\section{Conditions on Perfect Transfer}

Our primary goal is to prove necessary and sufficient conditions for perfect state transfer in the first excitation subspace of a spin-preserving Hamiltonian. These conditions can be expressed as the existence of a state transfer time $t_0$, and transfer phase $\phi$, in a condition on the eigenvectors
$$
|\braket{A}{\lambda_n}|=|\braket{B}{\lambda_n}|
$$
for all $n$, and on the eigenvalues,
$$
\lambda_nt_0=-\phi-\varphi_n+2\pi m_n
$$
for all $n$ for which $\braket{A}{\lambda_n}\neq 0$ where $m_n$ is an integer,
in close parallel to the equivalent results on chains \cite{mcpra,akreview}. However, we require the additional definition that 
$$
\varphi_n=\arg\left(\frac{\braket{\lambda_n}{B}}{\braket{\lambda_n}{A}}\right).
$$

To prove necessity, we start from the definition of state transfer in the single excitation subspace, requiring that there exists a $t_0$ and $\phi$ such that
\begin{equation}
e^{-iH_1t_0}\ket{A}=e^{i\phi}\ket{B}.		\label{eqn:definition}
\end{equation}
By taking the overlap with an eigenvector,
$$
e^{-i\lambda_n t_0}\braket{\lambda_n}{A}=e^{i\phi}\braket{\lambda_n}{B},
$$
one can immediately read off that $|\braket{A}{\lambda_n}|=|\braket{B}{\lambda_n}|$ by matching the weights. The phases must also match
$$
e^{-i\lambda_nt_0}=e^{i(\phi+\varphi_n)},
$$
up to a multiple of $2\pi$.

Having proved necessity, we prove sufficiency. Assume that a suitable $t_0$ and $\phi$ exist. So,
$$
e^{-iH_1t_0}\ket{A}=\sum_n\ket{\lambda_n}\braket{\lambda_n}{A}e^{-i\lambda_n t_0}.
$$
We can now supply the conditions on $\lambda_n$,
\begin{eqnarray*}
e^{-iH_1t_0}\ket{A}&=&\sum_{\braket{\lambda_n}{A}\neq0}\ket{\lambda_n}\braket{\lambda_n}{A}e^{i(2\pi m_n+\phi+\varphi_n)}	\\
&=&\sum_{\braket{\lambda_n}{A}\neq0}\ket{\lambda_n}\braket{\lambda_n}{A}e^{i\phi}\frac{\braket{\lambda_n}{B}}{\braket{\lambda_n}{A}}	\\
&=&e^{i\phi}\ket{B}.
\end{eqnarray*}

This yields a rather simple set of conditions which one use to verify that perfect transfer occurs in a network. Note that the task of testing the existence of a suitable time $t_0$ and phase $\phi$ can be largely neglected by taking differences and ratios of the eigenvalues.

One of the major advantages of such a characterisation for the spin chains was that this lead to the ability to calculate the required coupling strengths simply by specifying the desired spectrum. This was due to a specific property of spin chains in that, for mirror symmetric (real) Hamiltonians, after ordering the eigenvalues, $\varphi_n=\half(1+(-1)^n)\pi$ so it was very easy to specify a suitable spectrum. For real Hamiltonians, all the $\varphi_n$ must be either $0$ or $\pi$ (imposing that all the eigenvalues are integers, up to a scale factor and uniform shift), but as one varies the coupling strengths to adjust the eigenvalues, which $\varphi_n$ take which values can change. This was already observed as a practical problem in \cite{aklong}, making the task of designing perfect transfer networks far harder, though not impossible.

\section{Consequences}

With these necessary and sufficient conditions in place, we can start to explore the general features of transfer in networks. We want to know what is in principle possible without reference to specific Hamiltonians.

\subsection{Bipartite Graphs and the Transfer Phase}

Our first observation is a consequence of the study in \cite{pjpr}. For real Hamiltonians, constrained to a bipartite coupling graph (which also imposes that $B_n=0$), the transfer phase $e^{i\phi}$ is $\pm1$ if the transfer distance is even and $\pm i$ if the transfer distance is odd.

A bipartite graph is one whose vertices can be divided into two colourings, red and blue, such that edges only connect between a red and a blue vertex. Let us define
$$
S=\sum_{n\in\text{Red}}\proj{n}-\sum_{n\in\text{Blue}}\proj{n}.
$$
For a Hamiltonian $H_1$ which is connected via a bipartite coupling graph, it must be true that
$$
\{S,H_1\}=0.
$$
This means that for any eigenvector $\ket{\lambda_n}$ of $H_1$ with $\lambda_n\neq0$, $S\ket{\lambda_n}$ must also be an eigenvector of $H_1$, but with eigenvalue $-\lambda_n$. Now, let's assume (without loss of generality) that $A$ is in the `red' partition. We can write
$$
\ket{A}=\sum_{\stackrel{\lambda_n>0}{\braket{\lambda_n}{A}\neq 0}}\braket{\lambda_n}{A}(\ket{\lambda_n}+S\ket{\lambda_n})
$$
(for simplicity of notation, we have assumed that there are no 0 eigenvalues, but recall that we only need to consider one zero eigenvector with non-zero overlap on $A$, and it must satisfy $S\ket{\lambda_0}=\ket{\lambda_0}$, which allows us to treat this special case). Now let us evolve the state:
$$
e^{-iH_1t_0}\ket{A}=\sum_{\stackrel{\lambda_n>0}{\braket{\lambda_n}{A}\neq 0}}\braket{\lambda_n}{A}(e^{-i\lambda_nt_0}\ket{\lambda_n}+e^{i\lambda_nt_0}S\ket{\lambda_n},
$$
and calculate the overlap with some vertex $m$, remembering that for a real Hamiltonian the overlaps are real. If $m$ is a red vertex, $S\ket{m}=\ket{m}$, then
$$
\bra{m}e^{-iH_1t_0}\ket{A}=\sum_{\stackrel{\lambda_n>0}{\braket{\lambda_n}{A}\neq 0}}\braket{m}{\lambda_n}\braket{\lambda_n}{A}2\cos(\lambda_n t_0),
$$
so the amplitude is always real. Since $m$ was a red vertex, it must be an even distance from $A$. On the other hand, if $m$ is a blue vertex, then $S\ket{m}=-\ket{m}$ and
$$
\bra{m}e^{-iH_1t_0}\ket{A}=-\sum_{\stackrel{\lambda_n>0}{\braket{\lambda_n}{A}\neq 0}}\braket{m}{\lambda_n}\braket{\lambda_n}{A}2i\sin(\lambda_n t_0),
$$
so the amplitude is always imaginary. This provides another advantage when deciding if a network is capable of perfect transfer.

\subsection{Symmetries of the Hamiltonian} \label{sec:auto}

Symmetries are an important tool in understanding any system. Indeed, the construction of perfect state transfer chains originally relied heavily on an assumption of symmetry \cite{mcprl,mcpra}, which was subsequently \cite{akreview} proven to be necessary. We are thus interested in whether every perfect transfer Hamiltonian $H_1$ has a symmetry operator $S$ which satisfies $SH_1S^\dagger=H_1$ and $S\ket{A}=\ket{B}$.

The existence of a symmetry is proven by construction. By defining a unitary rotation that is diagonal in the basis of the Hamiltonian, it will clearly satisfy the commutation property. Specifying the phases as
$$
S=\sum_{\braket{A}{\lambda_n}\neq 0}e^{i\varphi_n}\proj{\lambda_n}+\sum_{\braket{A}{\lambda_n}=0}\proj{\lambda_n},
$$
allows us to verify the desired transformation
\begin{eqnarray*}
S\ket{A}&=&\sum_{\braket{A}{\lambda_n}\neq 0}e^{i\varphi_n}\ket{\lambda_n}\braket{\lambda_n}{A} \\
&=&\ket{B}.
\end{eqnarray*}
For a real Hamiltonian $H_1$, $S^2=\identity$, so $S\ket{B}=\ket{A}$. It is worth observing that there is still a continuous freedom in the definition of $S$ -- the phases that are applied to the eigenvectors for which $\braket{A}{\lambda_n}=0$ -- which gives a way to see that $S$ is not necessarily a permutation (which cannot be continuous). This manifests itself in the example of a chain below -- if $S$ were a permutation, it would have to be the mirror symmetry operator.

If one knows the symmetry operators of a system for some {\em a priori} reason, this identifies the values $\varphi_n$ (the eigenvalues of $S$), and associates them with specific eigenspaces. Hence, for systems where $S$ can be identified, and the eigenvalues can be modified while preserving the symmetry, we should be able to construct perfect transfer networks. This was the key insight for designing chains, and can hopefully now be applied in other scenarios.

We have mentioned at several points that a necessary condition on perfect end to end transfer chains is the presence of mirror symmetry \cite{akreview}. It is also the case that any Hamiltonian which achieves perfect transfer between opposite ends of a chain can equally achieve transfer between any mirror symmetric points. One might be drawn into the expectation that all perfect transfers (not just end to end) on chains are hence governed by mirror symmetric coupling schemes. This is not the case, as we will show by specific construction. Consider a matrix
$$
H_1=\left(\begin{array}{ccccc}
0 & J_1 & 0 & 0 & 0 \\
J_1 & 0 & J_2 & 0 & 0 \\
0 & J_2 & 0 & J_3 & 0 \\
0 & 0 & J_3 & 0 & J_4 \\
0 & 0 & 0 & J_4 & 0
\end{array}\right).
$$
One can prove that to transfer between qubits 2 and 4 (i.e.\ to create the evolution
$$
e^{-iH_1t_0}\ket{2}=e^{i\phi}\ket{4}
$$
up to some phase $\phi$, and for some time $t_0$), one simply has to impose that $J_1^2+J_2^2=J_3^2+J_4^2$ and that the eigenvalues of $H_1$ are (up to a scale factor) alternately even and odd integers. This eigenvalue condition is the same as for extremal transfer on the chain, but the coupling strengths are less restricted. You can readily verify that
$$
J_1=\sqrt{\frac{5}{2}-J_2^2}\qquad J_3=\frac{3}{2 J_2} \qquad J_4=\sqrt{\frac{5}{2}-\frac{9}{4 J_2^2}}
$$
is one class of non-symmetric examples which implement perfect transfer between qubits 2 and 4 with $t_0=\pi$.

\subsection{Transfer Rate}

Following on from some discussion of transfer rate in \cite{aklong}, \cite{akreview} examined the possibility of perfect transfer at high rate. This involves inserting a second state at site $A$ before the first state has been removed at site $B$, and yet requiring that the first state should still arrive perfectly. Some weak bounds were proven on possible rates for spin chains. We will now prove stronger bounds for all networks described by a real Hamiltonian (one can prove identical bounds for arbitrary Hamiltonians). A necessary condition for the ability to insert a second quantum state into the spin network (on the same input qubit) at some time $t$ without disturbing the first quantum state is that
$$
\bra{A}e^{-iH_1t}\ket{A}=0.
$$
For chains, this condition is sufficient, but for more general networks, this will not be the case. Ultimately, we will be interested in inserting many different states at different times. Again, for the chain, the only necessary condition is $\bra{A}e^{-iH_1t}\ket{A}=0$ for all of the possible time intervals $t$. For networks, the evolution of the many-excitation state could be quite different to the evolution of the single excitation states, so it might be that there are further conditions imposed. However, it is still a necessary condition, because our perfect transfer at high rate must work for all possible input states, which includes setting all previous inputs to $\ket{0}$ except for one, from which one can extract that same condition.

Thus, our question relates to whether, given there are $l$ unique time intervals $t_i<t_0$ at which $\bra{A}e^{-iH_1t_i}\ket{A}=0$, perfect transfer can occur to a site $\ket{B}$ at a distance $D$ in a time $t_0$. With $l$ time intervals, one can have $l$ unique times $t_i$ by imposing fixed intervals. We start by expressing our condition on the transfer distance as, for each integer $m=1\ldots D-1$,
$$
\bra{B}H^m\ket{A}=0.
$$
This can, instead, be written as
$$
\sum_{n=1}^Ne^{-i\varphi_n}\lambda_n^ma_n=0
$$
where $a_n=|\braket{A}{\lambda_n}|^2$, which is readily transformed into a linear equation
$$
\left(\sum_{m=0}^{D-1}\sum_{n=1}^Me^{-i\varphi_n}\lambda_n^m\ket{m}\bra{n}\right)\left(\sum_{n=1}^Ma_n\ket{n}\right)=0.
$$
Having resolved the possible degeneracies in the system, we have reduced from a system of size $N$ to $M$, the number of unique eigenvalues. Each of the $D-1$ rows is linearly independent.

The next constraint that we must add is that of normalisation,
$$
\left(\sum_{n=1}^M\bra{n}\right)\left(\sum_{n=1}^Ma_n\ket{n}\right)=1.
$$
Now we need to add in the conditions corresponding to $\bra{A}e^{-iH_1t_i}\ket{A}=0$. All our conditions so far have just been based on real values, and we will maintain this by dividing these conditions into real and imaginary parts. The real parts give
$$
\left(\sum_{i=1}^{l}\sum_{n=1}^M\cos(\lambda_nt_i)\ket{i}\bra{n}\right)\left(\sum_na_n\ket{n}\right)=0,
$$
and, similarly, the imaginary components give
$$
\left(\sum_{i=1}^{l}\sum_{n=1}^M\sin(\lambda_nt_i)\ket{i}\bra{n}\right)\left(\sum_na_n\ket{n}\right)=0.
$$
Given that all these times $t_i$ are less than $t_0$, the half period of the system (since we are assuming the Hamiltonian is real and performs perfect transfer, it is periodic with a period $2t_0$), all of these rows must be linearly independent from each other. Hence, if a suitable set of $a_n$ is to possibly exist, it must be the case that
\begin{equation}
2l+D\leq M\leq N.	\label{eqn:ratebound}
\end{equation}
In particular, imagine we had $M$ conditions not including the normalization condition. These would impose that all the $a_n=0$, so it would be impossible to satisfy the normalization condition. 

Ideally, we want the maximum transfer distance, which would be $N-1$ (a chain), imposing that $l=0$, as conjectured in \cite{akreview}. The only way to increase the perfect transfer rate is to reduce the transfer distance. However, you can't also lower the state transfer time (as you would expect by shortening the transfer distance). This is because the Margolus-Levitin theorem \cite{margolus} imposes a minimum time for evolving between two orthogonal states, such as a $\ket{1}_A$ as an input state, and the $\ket{0}_A$ required for the next input. Hence the transfer time is bounded from below by $(l+1)\pi/(4\sum_jJ_{1j})$.

In some sense, the `standard' perfect state transfer chains \cite{mcprl} saturate the bound of Eqn.\ (\ref{eqn:ratebound}) in that for a chain of $N$ qubits, any state $\ket{n}$ transfers a distance $D=N+1-2n$, but there are $n-1$ distinct times $t_i$ such that $\bra{n}e^{-iH_1t_i}\ket{n}=0$. Unfortunately, however, these times are not equally spaced, so they are not useful for achieving a high rate of transfer. It is worth noting that our analysis breaks down at the $l=0$ limit since $t_0$ is the length of the period, not the half period. This means that half of the $2l$ conditions can be the same as the other half, for suitably chosen values of $t_i$. We end up with $l\leq N$, and this bound was saturated in \cite{akreview} for the sequential quantum storage solution.


\subsection{Routing}

The idea of being able to choose which of several recipients, $B$, $C$ etc., is to receive a quantum state was initially studied in \cite{arbac}, and some aspects have been further considered in \cite{nik,nik2}\footnote{A word of warning is warranted, however. Due to the non-uniqueness of a non-integer power of a unitary, $U^k$, there may be gaps in some of the proofs, such as Theorem 1 in \cite{nik2}.}. This task has since become known as routing \cite{pjprprl} and, by allowing some minimal control, it was achieved efficiently in a regular network of nearest-neighbour coupled spins. Routing is potentially an important property for a system to possess since this allows us to significantly alter the connectivity of an array of sites in a way that direct communication between pairs of sites does not. However, we are now going to make a proof by contradiction that shows that for real Hamiltonians, routing between multiple sites is impossible, and subsequently we will bound the number of possible recipients as a function of transfer distance for more general Hamiltonians. This is something that the constructions of, for instance \cite{nik}, give no information about, or control over. While they allow for the inversion from a desired unitary to a Hamiltonian, this provides no control over any spatial limitations in the coupling patterns, variations of coupling strengths (such as uniformity of coupling, or even real values) and typical solutions couple every qubit to every other qubit in a completely arbitrary manner, entirely missing the point of state transfer, which is intended for use in systems of low connectivity.

We start by assuming that perfect transfer is possible between $A$ and $B$, and the minimum time in which this occurs is $t_{AB}$. So, we have
$$
e^{-iH_1t_{AB}}\ket{A}=e^{i\phi}\ket{B}.
$$
Recall that since the Hamiltonian is real, all the $\varphi_n$ are 0 or $\pi$. So, this means that if we evolve for twice the time, we have a perfect revival,
$$
e^{-i2H_1t_{AB}}\ket{A}=e^{2i\phi}\ket{A},
$$
demonstrating that the dynamics are periodic. Now let us assume that perfect routing is possible, meaning that there must exist a time $t_{AC}<t_{AB}$ such that
$$
e^{-iH_1t_{AC}}\ket{A}=e^{i\phi'}\ket{C}.
$$
However, by identical arguments, it must be the case that
$$
e^{-2iH_1t_{AC}}\ket{A}=e^{2i\phi'}\ket{A}
$$
and hence
$$
e^{-iH_1(2t_{AC}-t_{AB})}\ket{B}=e^{i(2\phi'-\phi)}\ket{A}.
$$
This is just perfect transfer between $B$ and $A$ in time $|2t_{AC}-t_{AB}|<t_{AB}$, which is impossible by assumption that $t_{AB}$ is the shortest state transfer time. Hence the transfer from $A$ to $C$ can't exist, and if there is transfer to one site, there cannot be transfer to any other sites. In order to break this restriction, we have to take Hamiltonians with complex entries. This is exactly what happens in papers such as \cite{nik}.

The preceding argument is rather powerful, revealing that many other intermediate states can't exist, since once can repeat it for any target state which is just a superposition of eigenvectors where, up to a global phase, all the amplitudes are real. One obvious example comes from the bipartite systems we discussed previously -- a real Hamiltonian on a bipartite lattice which is capable of perfect transfer can never, at any intermediate time, produce a state which is entirely localised on just one of the bipartitions of the graph.

In the previous subsection, we derived a trade-off between the maximum transfer rate and the distance of transfer. We can do the same for a general case of routing, where we wish to transfer to $J$ different possible sites from $A$, at locations $j$ and times $t_j$. As before, we have $a_n=|\braket{A}{\lambda_n}|^2$ and $\braket{A}{\lambda_n}=e^{i\varphi_{n,j}}\braket{j}{\lambda_n}$. Now we have $J$ conditions for the perfect transfer
$$
\left(\sum_{j=1}^J\sum_{n=1}^N\ket{j}\bra{n}e^{-i\lambda_nt_j}\right)\left(\sum_na_n\ket{n}\right)=0
$$
(the independence of these conditions is no longer imposed by periodicity, of which we are not assured, but by the assumption that the output vertices are distinct, yielding orthogonal states) and there's the same normalisation condition. Now we also want to impose that all $J$ target vertices are at least a distance $D$ from $A$. Hence, for $k=1$ to $D-1$ we have
$$
\left(\sum_{j=1}^J\sum_{n=1}^N\ket{j}\bra{n}\lambda_n^ke^{i\varphi_{n,j}}\right)\left(\sum_na_n\ket{n}\right)=0.
$$
Even by not restricting the $a_n$ to be real (let alone positive), we arrive at the bound
$$
DJ\leq M-1\leq N-1.
$$
So, if you want to route between every possible vertex of a network, you must have transfer distance 1. This is exactly what happened in the examples of \cite{nik}, but now we know that it's impossible to do better, making the results of \cite{pjprprl} all the more remarkable, achieving routing at a high transfer rate with only the addition of very modest controls.

The interpretation of the periodicity of the system also allows a minor insight into the state transfer time. Let us define the eigenvalue gaps between eigenvectors with support on $A$ as $\Delta_n=\lambda_{n+1}-\lambda_n$ (where the $\lambda_n$ are ordered), and fix $\chi$ to be the largest real number such that $\Delta_n/\chi$ is an integer for all $n$. Then it must be the case that the state transfer time is given by $t_0=\pi/\chi$. This is because it allows $e^{i\Delta_n\pi/\chi}$ to be $\pm1$, as would be required for state transfer, and yet $e^{2i\Delta_n\pi/\chi}=1$, which corresponds to a perfect revival on the input spin.

\subsection{Uniformly Coupled Systems} \label{sec:uniform}

Perhaps of most interest would be finding graphs which are uniformly coupled (to be defined momentarily), preferably maximising the growth of transfer distance with the total number of vertices, and keeping the degree of each vertex low.

There are two natural connections between Hamiltonians restricted to the single excitation subspace and the underlying graph structure. The first is the $XX$ model,
$$
H=\half\sum_{\{i,j\}\in E}X_iX_j+Y_iY_j,
$$
which has $H_1=A$, the adjacency matrix of the graph $G$ with edge set $E$. The second is the Heisenberg model,
$$
H=-\half\sum_{\{i,j\}\in E}X_iX_j+Y_iY_j+Z_iZ_j+\identity,
$$
which has $H_1=L$, the graph Laplacian. 

\subsubsection{Heisenberg/Laplacian Systems}

Our strategy for the two cases will be slightly different, and we start with the Heisenberg case. From our necessary and sufficient conditions for state transfer, we know that the eigenvalues of $H_1$ can be written in the form
$$
\lambda_n=\chi z_n+\delta
$$
for the eigenvectors $\ket{\lambda_n}$ which have support on the input vertex, where $z_n$ is a different integer for each $n$, but $\chi$ and $\delta$ are fixed (and relate to the transfer time and phase respectively). In fact, $\delta=0$ because we know that a Laplacian always has one eigenvector
\begin{equation}
\ket{\lambda}=\frac{1}{\sqrt{N}}\sum_{n=1}^N\ket{n}	 \label{eqn:maxev}
\end{equation}
with eigenvalue 0.
We are now going to assume that every unique eigenspace has support on the input vertex. Under this assumption, we can calculate
$$
\Tr(H_1)=\chi\sum_{n=1}^Nz_n=\sum_nd_n
$$
where $d_n$ is the degree of vertex $n$. This instantly proves that $\chi$ is rational, and hence all the eigenvalues are rational. However, it is well known \cite{graph_theory} that any Hamiltonian with integer matrix elements and rational eigenvalues in fact has integral eigenvalues. Hence, we can utilise the wide variety of results on Laplacian integral graphs \cite{lap_int}. Nevertheless, it is worth emphasising that it is only necessary that the graphs be integral. It is in no way sufficient.

Now we want to know about how the transfer distance is related to the maximum degree of the graph and the number of vertices, under this assumption about the support of the eigenvectors. For a connected graph, the diameter $D$ (the maximum distance between any two points in the graph, which is an upper bound on the transfer distance) is bounded by $k$, the number of distinct eigenvalues: $D+1\leq k$ \cite{laplacian}. However, since the minimum eigenvalue is 0, and they are spaced by integers, the maximum eigenvalue must be larger than $k-1$, and yet is upper bounded by $2d$, where $d$ is the maximum degree of any vertex. We conclude that
$$
D\leq 2d.
$$
Unfortunately, to have a scaling transfer distance, we must scale the maximum degree of the graph. More general bounds are stated in \cite{mohar}, such that for any $\alpha>1$,
$$
D\leq 2\left\lceil\sqrt{2d}\sqrt{\frac{\alpha^2-1}{4\alpha}}+1\right\rceil\lceil\log_{\alpha}(N/2)\rceil.
$$
This suggests that perhaps the logarithmic trade-off between transfer distance and number of qubits in the hypercube \cite{mcpra} might be necessary. However, we are still far from proving this. Instead, we only have the upper bound on $D$ of $O(\sqrt{d}\log N)$.

\subsubsection{\texorpdfstring{$XX$/Adjacency Systems}{XX/Adjacency Systems}}

For the $XX$ model, any eigenvectors with support on the input vertex have eigenvalues
$$
\lambda_n=\chi z_n+\delta,
$$
where $z_n$ is a rational number. Again, we will assume that all eigenspaces have support on $A$. This is equivalent to the assumption made in \cite{godsil}, which imposed that every vertex in the graph be periodic (i.e.\ have a perfect revival after $2t_0$), and, indeed, we will arrive at the same conclusions. We also note that there are known instances where perfect transfer can be found without needing this assumption \cite{angeles}. As before, we proceed by calculating
$$
\chi\sum_nz_{n=1}^N+N\delta=\Tr(H_1)=0.
$$
Hence, $\delta$ is a rational number multiplied by $\chi$ (thereby imposing that the transfer phase is a root of unity), which we incorporate into $z_n$, so $\lambda_n=\chi z_n$. Next consider the characteristic polynomial of $H_1$,
$$
\text{det}(H_1-\lambda\identity)=\sum_{n=0}^Na_n\lambda^n=0,
$$
which contains integer coefficients $a_n$ (because all the matrix elements of $H_1$ are integers).
Each of the $a_n$ can be equated with a combination of the $(N-n)^{th}$ order products of the eigenvalues, which therefore appear as $\chi^{N-n}$ multiplied by a rational number, so $\chi^{N-n}$ is rational. It is always true that $a_{N-2}\neq 0$, so $\chi^2$ is rational. Provided the graph is not bipartite, there is always a value of $k$ for which $a_{N-2k-1}\neq 0$, so $\chi^{2k+1}$ is rational, and hence $\chi$ is rational. Again, this imposes that, in fact, the graph is integral and we can use the many results on integral graphs \cite{integral}. However, if the graph is bipartite, these conclusions do not hold. The simplest counter-example is the chain of 3 qubits which has eigenvalues $0,\pm\sqrt{2}$ and yet achieves perfect transfer.

For these systems, we are not aware of any general bounds trading between the degree and transfer distance of the graph. However, it is worth noting that if the graph is regular (i.e.\ every vertex has the same degree, $d$), the Laplacian, $L$, and adjacency matrix $A$ are related by
$
L=d\identity-A
$
and hence the previous bounds apply.

\begin{figure}
\begin{center}
\includegraphics[width=0.45\textwidth]{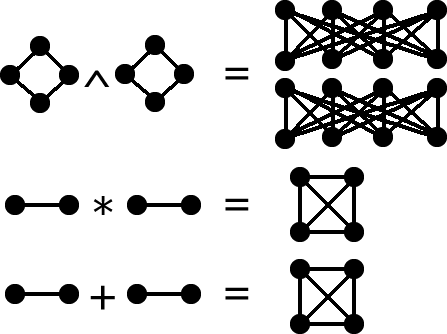}
\end{center}
\vspace{-0.5cm}
\caption{Simple examples of the conjunction, strong product and join of two perfect state transfer graphs which do not produce perfect state transfer graphs. This can be determined by inspection -- the graphs are too symmetric to single out a target vertex for any given input vertex.} \label{fig:counter}
\end{figure}

In \cite{integral}, several ways of combining and manipulating integral graphs to give new integral graphs were proven:
\begin{description}
\item[Cartesian product of two integral graphs] \hfill \\
$G=G_1\times G_2$ has vertices $V=V_1\times V_2$ with edges between $\{(u_1,u_2),(v_1,v_2)\}$ if $\{u_1,v_1\}\in E_1$ or $\{u_2,v_2\}\in E_2$. This is the exclusive or.
\item[Conjunction of two integral graphs] \hfill \\
$G=G_1\wedge G_2$ has vertices $V=V_1\times V_2$ with edges between $\{(u_1,u_2),(v_1,v_2)\}$ if $\{u_1,v_1\}\in E_1$ and $\{u_2,v_2\}\in E_2$ (also known as the tensor product of two graphs).
\item[Strong product of two integral graphs] \hfill \\
$G=G_1*G_2$ has vertices $V=V_1\times V_2$ with edges between $\{(u_1,u_2),(v_1,v_2)\}$ if $\{u_1,v_1\}\in E_1$ or $\{u_2,v_2\}\in E_2$ or both.
\item[Join of two regular integral graphs] \hfill \\
$G=G_1+G_2$ has vertices $V=V_1+V_2$ with edges $\{u,v\}\in E$ if $\{u,v\}\in E_1$ or $\{u,v\}\in E_2$ or if one vertex is part of each graph, with the restriction that $(d_1-d_2)^2+4N_1N_2$ must be a perfect square.
\item[Complement of an integral graph] \hfill \\
$G$ is the same as the original graph but the edge set is inverted.
\end{description}
One might therefore wonder if these same constructions can take perfect transfer graphs (with the same transfer time) and produce new perfect transfer graphs. This study started in \cite{mcpra} which showed how, by taking two graphs known to exhibit perfect transfer, one can construct a larger graph, via the graph product, that also exhibits perfect transfer. While this was shown specifically for chains, it is easily generalised to all perfect transfer graphs \cite{angeles}. By way of contrast, we give examples in Fig.\ \ref{fig:counter} for which the conjunction, strong product and join do not generate perfect transfer graphs\footnote{While one could show that the conjunction of any bipartite graph with either the two or three vertex chain, this is a trivial result since both just produce two independent copies of the original graph.}.

The complement is a more interesting case. In general, it cannot be true that there is perfect transfer, which one sees by considering the graph of two connected vertices. This exhibits perfect transfer, but the complement, which is the unconnected graph, clearly does not perform transfer. However, there are some cases where the complement does give a perfect transfer graph, and these instances are readily verified. Let $\ket{\lambda}$ be the state specified in Eqn.\ (\ref{eqn:maxev}). Since the graph is regular, $\ket{\lambda}$ is the maximum eigenvector, with eigenvalue $d$. We can use this to write the complement $\bar A$ of the adjacency matrix,
$$
\bar A=N\proj{\lambda}-\identity-A.
$$
So, if we perform a state transfer between the same input and output vertices in the same time $t_0$, since $\proj{\lambda}$ commutes with $A$, the condition on achieving state transfer with the complement is simply
$e^{-it_0N}=1$,
which also applies to the Laplacian of a regular graph.

\section{Conclusions}

In this paper, we have given necessary and sufficient conditions for the existence of perfect state transfer in a quantum network, using the single excitation subspace. One should be aware, however, that the dynamics can be much richer in higher excitation subspaces, with the possibility of catalysing otherwise impossible transfers \cite{pjpr}.

Making use of these conditions allows us to easily decide if a system can perform perfect state transfer. We have proven a bound on the maximum transfer rate. The routing of quantum states between multiple different sites is impossible if the Hamiltonian is real, although these results do not contradict existing schemes for the arbitrarily accurate scenario \cite{arbac}, when some degree of control is allowed \cite{pjprprl}, or when complex coupling coefficients are allowed \cite{nik,nik2}.

We hope that the insights provided in this paper lead to progress in designing perfect state transfer Hamiltonians in a wider class of systems. Using results from spectral graph theory, we have already been able to place bounds on many of the properties of uniformly coupled networks, such as the bound that the transfer distance can be no more than twice the degree of the graph for the Heisenberg model or regular $XX$ models (provided the graph is periodic), and anticipate that much more is possible. We also think it is important that the vague suggestion that networks should be more robust to perturbations should be put on a more rigorous footing.

\acknowledgments
This work is supported by the National Research Foundation \& Ministry of Education, Singapore.

\end{document}